\documentclass{osa-article}
\usepackage{siunitx}
\DeclareSIUnit{\belmilliwatt}{Bm}
\DeclareSIUnit{\dBm}{\deci\belmilliwatt}
\usepackage{enumitem}

\journal{oe}
\newcommand{\CMT}[1]{{}}

\articletype{Research Article}
\begin{document}

\title{Mitigating photorefractive effect in thin-film lithium niobate microring resonators}

\author{Yuntao Xu\authormark{1},  Mohan Shen\authormark{1}, Juanjuan Lu\authormark{1}, Joshua Surya\authormark{1}, Ayed Al Sayem\authormark{1}, and Hong X. Tang\authormark{1,*}}

\address{\authormark{1}Department of Electrical Engineering, Yale University, New Haven, Connecticut 06520, USA}

\email{\authormark{*}Corresponding author: hong.tang@yale.edu} %% email address is required% \homepage{http:...} %% author's URL, if desired

%%%%%%%%%%%%%%%%%%% abstract %%%%%%%%%%%%%%%%
%% [use \begin{abstract*}...\end{abstract*} if exempt from copyright]

\begin{abstract}
Thin-film lithium niobate is an attractive integrated photonics platform due to its low optical loss and favorable optical nonlinear and electro-optic properties. However, in applications such as second harmonic generation, frequency comb generation, and microwave-to-optics conversion, the device performance is strongly impeded by the photorefractive effect inherent in thin-films. In this paper, we show that the dielectric cladding on lithium niobate microring resonator has a significant influence on the photorefractive effect. By removing the dielectric cladding layer, the photorefractive effect in lithium niobate ring resonators can be effectively mitigated. Our work presents a reliable approach to control the photorefractive effect on thin-film lithium niobate and will further advance the performance of integrated classical and quantum photonic devices based on thin-film lithium niobate.
\end{abstract}

%%%%%%%%%%%%%%%%%%%%%%%%%%  body  %%%%%%%%%%%%%%%%%%%%%%%%%%

\section{Introduction}

As one of the most widely used synthetic crystals, lithium niobate (LN) has played a critical role in modern telecommunication due to its favorable electro-optic and optical $\chi^{(2)}$ nonlinearity as well as broad optical transparency window \cite{weis1985lithium,boyd2019nonlinear,garcia2014imaging,leidinger2015comparative}. Traditional optical waveguides on LN are fabricated using titanium in-diffusion or proton exchange method \cite{wooten2000review,parameswaran2002highly}, which leads to weak optical confinement due to the large mode volume. With recent development on nanofabrication technique, low-loss waveguides and high-quality microresonator have been demonstrated on monolithic thin-film lithium niobate on insulator (LNOI) material system \cite{zhang2017monolithic}, providing a promising on-chip platform for second harmonic generation (SHG) \cite{lu2019periodically,wang2017second}, frequency comb generation \cite{gong2019soliton,he2019self,wang2019monolithic}, and electro-optic modulation \cite{wang2018integrated}.

In many of these applications, the performance and power handling of devices fabricated on the LN platform have been strongly limited by photorefractive (PR) effect \cite{lu2019periodically,kong2020recent}.  Despite intensive studies on PR effect in both doped and intrinsic bulk LN crystals \cite{bryan1984increased,volk1994optical,razzari2005photorefractivity,von1978intrinsic,ludtke2011light}, there remains a considerable knowledge gap between the study of PR effect in thin-film LN and that of bulk LN \cite{liang2017high,li2019high,li2019photon}. Relaxation time of $\sim$ \SI{20}{\milli\second} in suspended LN microdisk cavity has been observed \cite{liang2017high}, significantly faster than the relaxation process reported in bulk LN \cite{maniloff1993effects}. While quenching of PR effect has been reported in both photonic crystal cavity and microring resonator \cite{li2019high,wang2019monolithic}, the underlying mechanism remains subject to further investigation.  

In this paper, we study the impact of dielectric cladding and the following heat treatment on the PR effect of thin-film LNOI microring resonators. Cladding the waveguides and rings with dielectric is a standard fabrication process to protect optical structures and engineer the dispersion. However, we find that the commonly employed oxide cladding dramatically exacerbates the PR effect of the thin-film LN ring resonator. By removing the silicon dioxide cladding and applying proper heat treatment, the relative resonance frequency shift induced by the PR effect could be suppressed by more than 
two orders of magnitude with the same intra-cavity power, suggesting a strong correlation between the PR effect of thin-film LN devices and its interface condition. These results not only provide a guideline for the community on how to mitigate and engineer the strength of PR effect but also potentially motivate cross-disciplinary researches on the mechanism of the PR effect on thin-film LN. 

\begin{figure}[h]
    \centering
    \includegraphics[width=0.8\textwidth]{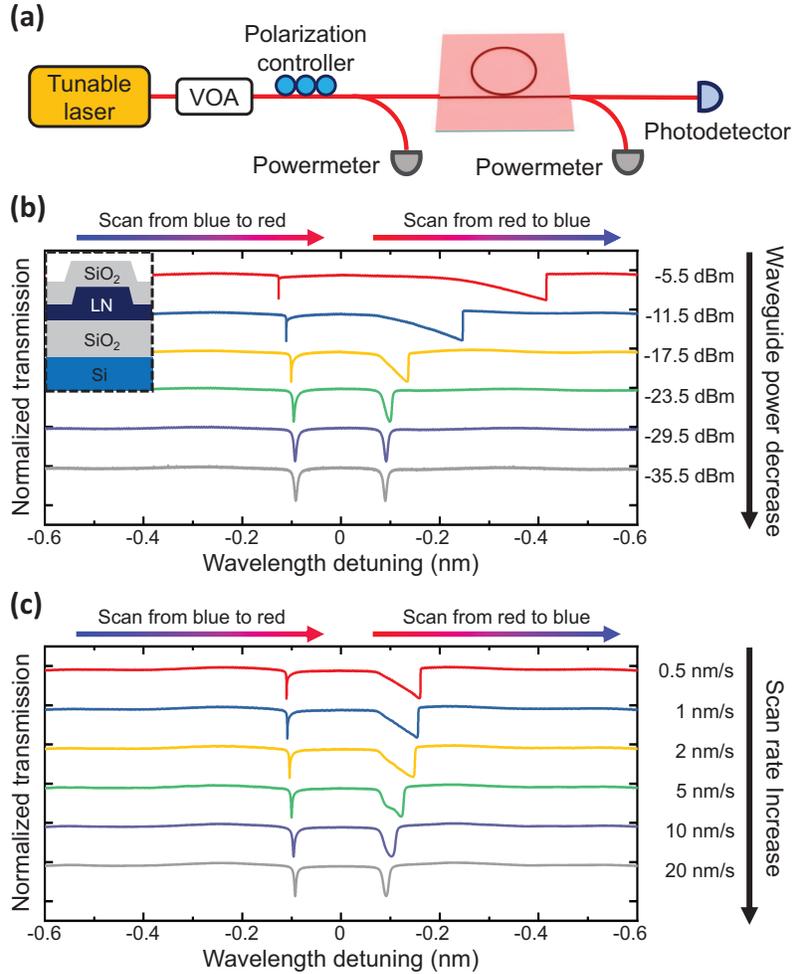}
    \caption{(a) Experimental setup to characterize the PR effect in LN microresonators. Surface grating couplers are exploited for fiber-to-chip coupling. VOA: variable optical attenuator. (b) Forward and backward laser-scanned transmission spectra of a PECVD oxide-cladded LN ring resonator under different optical power with a fixed scan rate \SI{0.5}{\nano\metre/\second}. The power in the waveguide varies from \SI{-5.5}{\dBm} to \SI{-35.5}{\dBm}. 
   The inset shows the cross-sectional sketch of the oxide-cladded waveguide. (c) Forward and backward transmission spectra of the same LN ring resonator with different scan rates at a fixed waveguide power of \SI{-15.5}{\dBm}. The scan rate varies from \SI{0.5}{\nano\metre/\second} to \SI{20}{\nano\metre/\second}.
   % \question{( 1. place the panel labels a,b,c,d,e to upper left of each panel. 2. label absolute power in each curve in (b,c) and the sweep directions. 3. Add space between panels (b,c) and (d,e). 4. Consider put the trace-retrace curves in a single panel, for example the red traces in b,c could be combined. This will probably give you a better view. 5. The goal is to have your reader to understand the curves without reading your caption.)}
    }
    \label{fig_setup}
    \end{figure}

\section{PR effect in the LN microring resonator}

Our first batch of devices are fabricated from \SI{600}{\nano\meter} x-cut LN on \SI{2}{\micro\metre} thermal-oxide intermediate layer on high-resistivity silicon wafer. The optical ring resonators have a radius of \SI{80}{\micro\metre} and a width of \SI{1.6}{\micro\metre}, patterned by electron beam lithography (EBL) with hydrogen silsesquioxane (HSQ) resist. Subsequently, the pattern is transferred to thin-film LN using $\mathrm{Ar^+}$-based inductively coupled plasma (ICP) reactive ion etching (RIE), with \SI{350}{\nano\metre} LN being etched. The device is then dipped in buffered oxide etchant (BOE) to remove the resist residue and coated with \SI{600}{\nano\metre} amorphous silicon dioxide with plasma enhanced chemical vapor deposition (PECVD) as a standard cladding layer \cite{gondarenko2009high,zhang2017monolithic,kippenberg2018dissipative,riemensberger2012dispersion}. An one-hour annealing process at \SI{500}{\degreeCelsius} in a nitrogen environment is followed to improve the quality of silicon dioxide and reduce the optical loss of the ring resonators. 

The schematic diagram of the experimental setup to characterize the PR effect of the LN microring resonator is shown in Fig.\,\ref{fig_setup}(a). Light from a telecom tunable laser (Santec-710) is coupled to the waveguide through a grating coupler, while the input power is controlled by the variable optical attenuator (VOA) and monitored by an optical power meter. The device under test is placed on an XYZ motorized stage to tune the fiber-to-chip insertion efficiency. During the measurement, the optical power input to grating coupler is \SI{4.5}{\dBm} when the attenuation of VOA reads \SI{0}{\decibel}. The attenuation of VOA and scan rate of the tunable laser are varied from \SI{0} to \SI{30}{\decibel}, and \SI{0.5} to \SI{20}{\nano\metre/\second}, respectively. The total insertion loss of a grating coupler pair is tuned to $1.0\%$, and thereby the coupling efficiency is $10\%$ each facet due to device symmetry.

%\question{(Did you actually tune the grating coupler efficiency? It is a bit rough to say there is 10dB loss at each facet. There should be one extra significant digit.)} 

A fundamental transverse-electric (TE) mode at \SI{1557.61}{\nano\metre} with an optical quality factor of $2.0 \times 10^5$ is selected to investigate the PR effect. The forward (from short to long wavelength) and backward (from long to short wavelength) laser-scans across the resonance are applied, whose transmission spectra are recorded by normalizing the photodetector response to the reference power meter, as presented in Fig.\,\ref{fig_setup}(b). 

For the device with PECVD oxide-cladding, a strong PR effect is observed. A series of spectra of the selected mode are taken at different input powers and scan rates. % to analyze the PR effect in the PECVD oxide-cladding ring resonator. 
As shown in Fig.\,\ref{fig_setup}(b), when the scan rate is fixed at \SI{0.5}{\nano\metre/\second}, the resonance shape is significantly deformed from an ideal Lorentzian shape as the waveguide power increases from \SI{-35.5}{\dBm} to \SI{-5.5}{\dBm}. When the laser scans forward (backward), the resonance shape experiences severe compression (broadening), indicating the resonance frequency shift to a shorter wavelength as laser frequency approaching the resonance frequency. This behavior confirms that the PR effect occurs in the microring cavity and induces a decrease of refractive index. We then fix the on-chip waveguide power at \SI{-15.5}{\dBm} and vary the laser scan rate from \SI{0.5}{\nano\metre/\second} to \SI{20}{\nano\metre/\second}. The forward and backward scan traces are respectively shown in Fig.\,\ref{fig_setup}(c), where the resonance shape evolves from a Lorentzian-like shape to a triangle shape when scanned from blue to red as the scan rate decreases. Such scan rate dependence indicates a characteristic time of 
$\SI{10}{\milli\second}$ scale, which is consistent with the value reported in the LN microdisk resonator \cite{liang2017high,jiang2017fast}. The detailed data analysis and fitting will be described in section\,\ref{sec:scandynamic}.

\begin{figure}[t]
    \centering
    \includegraphics[width=0.5\textwidth]{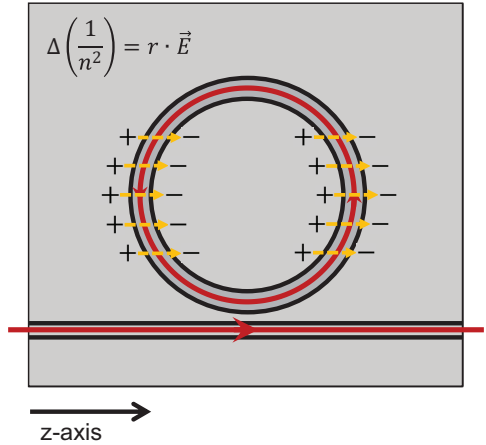}
    \caption{Schematic illustration of the PR effect in x-cut thin-film LN ring resonator. The space-charge field builds up due to the asymmetric excitation of the light-induced carriers in the LN ring resonator, which subsequently modulate the refractive index of the ring resonator. %\question{This figure is kind of too simple. You don't really need panel b. Your placement of $+$ and $-$ across the waveguide is kind of arbitrary. Maybe you can move them closer to align with the edge of the waveguide. Keep panel (a) only. Move equation in (b) to (a)}
    }
    \label{fig_mech}
\end{figure}

\section{Mitigation of the PR effect by removing the cladding layer}

The PR effect in LN induces a resonance frequency shift that produces unfavorable instability in device applications \cite{lu2019periodically,liang2017high}. It arises as a combination of several effects that induce a refractive index change under optical illumination \cite{hall1985photorefractive}. When the LN crystal is illuminated, the photo-excited carrier has a higher possibility of being guided along certain directions due to the broken inversion symmetry of the crystal. These charges then migrate to unoccupied trapping regions in the crystal and induce a charge separation in the material --- an effect known as the bulk photovoltaic effect. The macroscopic manifestation of this microscopic behavior is a space-charge field when the material is illuminated \cite{weis1985lithium}. The space-charge field subsequently modulates the refractive index of the LN crystal through the linear electro-optic Pockels effect.

The PR effect observed in thin-film LN is significantly different from that in bulk LN \cite{kong2020recent,liang2017high,jiang2017fast,li2019high}. A conceptional illustration of the PR effect in thin-film LN ring resonator is shown in Fig.\,\ref{fig_mech}. Comparing to the bulk crystal, interface and surface states of the material aggravate the PR effect on the thin films. The impact of interface state could be potentially attributed to two channels. First, the interface state at the LN-oxide interface can provide extra donor and trapping sites for the carrier excitation and migration process \cite{kanemitsu1990photocarrier,mandelis2005theory}. Second, extra second-order nonlinearity could arise from surface dipole effect induced by broken interface symmetry \cite{levy2011harmonic,bloembergen1968optical}. In our waveguide structure, two interfaces could be responsible for the observed strong PR effect: the top PECVD oxide cladding interface and the underneath LN-oxide bonding interface. Since the latter is beyond our control for the unsuspended structures, we could remove the top PECVD silicon dioxide layer to examine the impact of the LN-PECVD oxide interface. 

\begin{figure}[b]
    \centering
    \includegraphics[width=0.8\textwidth]{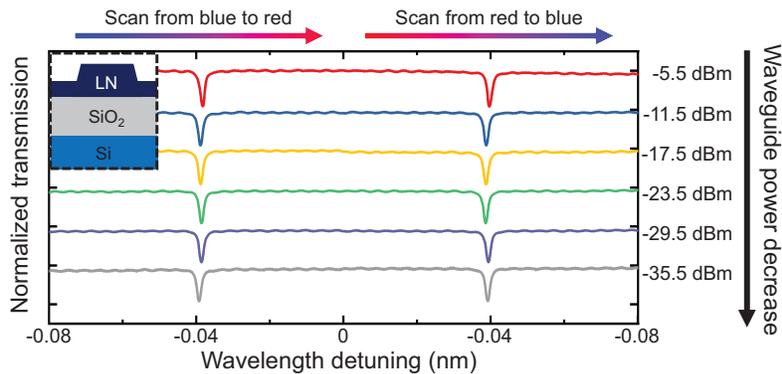}
    \caption{Forward- and backward-scanned transmission spectra of an air-cladded x-cut LN ring resonator at a fixed scan rate of \SI{0.5}{\nano\metre/\second}. The on-chip waveguide power varies from \SI{-5.5}{\dBm} to \SI{-35.5}{\dBm}. The inset shows the cross-section of air-cladded waveguides.}
    \label{fig_air}
\end{figure}

We then fabricated the second batch of devices without PECVD oxide cladding. After plasma etching and removal of residual HSQ resist, the air-cladded devices are directly annealed at \SI{500}{\degreeCelsius} in the nitrogen environment for \SI{1}{\hour}. As shown in Fig.\,\ref{fig_air}, distinct to the oxide-cladded device, with the same pump power as that of the previous measurement in Fig.\,\ref{fig_mech}(b), no resonance shape deformation could be observed in the air-cladded device, indicating that oxide cladding exacerbates the PR effect and the optimized air-cladding fabrication flow dramatically mitigates the PR effect.  

\begin{figure}
    \centering
    \includegraphics[width=0.6\textwidth]{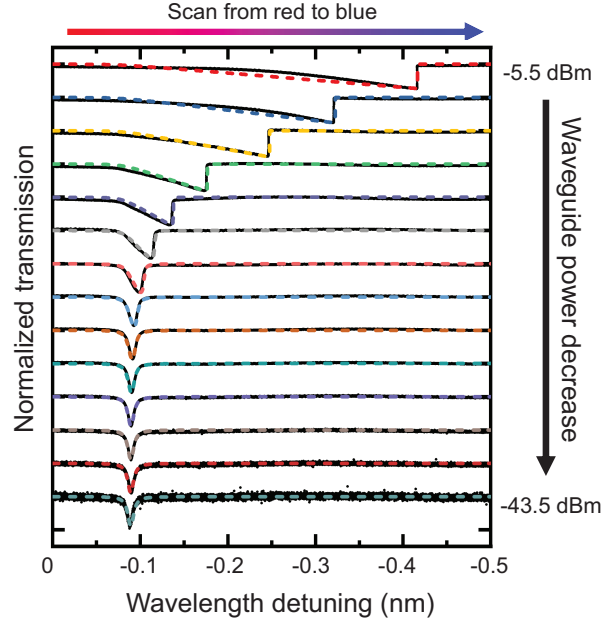}
    \caption{ Backward laser-scanned transmission spectra of an x-cut LN ring resonator cladded with PECVD oxide. The black dots are experiment data, and colored dashed lines are corresponding fitting curves. The scan rate is fixed at \SI{0.5}{\nano\meter/\second} and the waveguide power is varied from \SI{-35.5}{\dBm} to \SI{-5.5}{\dBm}. %\question{(Figure very difficult to interpret. I don't read any fittings.  I propose you do the following 1) use one single panel; 2) all curves use a single color (say, black dots); 3) fit with a blue short-dashed trace; fitting overlaid with experimental traces. 4) I don't like your image-like figures. They are not professional and do not reflect well on us. They also make files unnecessarily large. They should be all vector plots, not jpg or bmp or png)}
    }
\label{fig_fitting}
\end{figure}

\section{Cavity dynamics in the LN ring resonator during the laser scan}
\label{sec:scandynamic}

The cavity dynamics in the ring resonator can be described by a model that incorporates both photothermal \cite{carmon2004dynamical} and PR \cite{sun2017nonlinear} effects. In this study, due to the use of a relatively weaker probe power as well as the favorable thermal conductivity compared to previously studied suspended structures, the cavity dynamics are dominated by the PR effect with negligible thermo-optical effect. The simplified equation of motion of cavity dynamics could be written as

\begin{equation}
\frac{da}{dt}=-(i\delta+\frac{\kappa}{2})a-i g_E E_{\rm{sp}} a + \sqrt{\kappa_{\rm{ex}}} a_{\rm{in}, } \label{Eq_Motiona}
\end{equation}

\noindent Here $a$ is the annihilation operator for the optical resonance mode, and $\delta=\omega_0-\omega_p$ is the pump detuning. $\kappa$ and $\kappa_{\rm{ex}}$ are the total decay rate and external decay rate of the mode $a$, respectively. The term $g_E E_{\rm{sp}}$ represent resonance frequency shift due to the PR effect, where $E_{\rm{sp}}$ is the average space-charge induced electric field weighted over the optical mode profile and $g_E = \frac{n_0^2}{4} \omega_0 {(r_{13}+r_{33})}$ is electro-optic coupling coefficient contributed by the Pockels effect. $a_{\rm{in}}$ denotes the input waveguide mode and $P_{\rm{pump}}=\hbar \omega_p |	\langle a_{\rm{in}} \rangle |^2$ is the pump power in waveguide.

The space-charge field is generated via photon absorption and followed charge migration, thus the dynamic of the space-charge field can be described by \cite{sun2017nonlinear}

\begin{equation}
\frac{dE_{\rm {sp}}}{dt} = -\alpha_E E_{\rm {sp}} + \eta_E |\langle a \rangle|^2 \label{Eq_Esp}
\end{equation}

\noindent where $\alpha_E$ and $ \eta_E$ are respectively the relaxation rate and the generation coefficient of the space-charge field. Here we assume that $\eta_E$ is independent to optical power.

Equations (\ref{Eq_Motiona}, \ref{Eq_Esp}) completely describe the intracavity dynamics in the optical cavity under the PR effect, through which we can fit the unknown factors $\alpha_E$ and $\eta_E$ from the recorded transmission spectra. The fitted curve is shown in Fig.\,\ref{fig_fitting}. We note that as the optical power increases, the fitting curve slightly deviates from the experimental data, suggesting nonlinearity of space-charge field generation at high optical powers. Although the understanding underlying mechanism for this saturation behavior is subject to future studies, one possible reason is the power-dependent charge excitation and trapping efficiency \cite{ludtke2011light}.

\begin{figure}[b]
    \centering
    \includegraphics[width=0.7\textwidth]{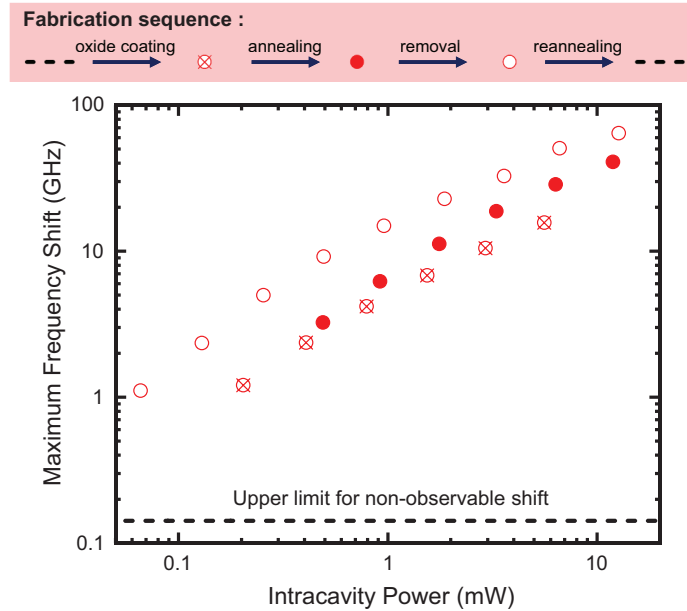}
    \caption{ The PR-induced maximum frequency shift in LN ring resonators processed under different conditions as a function of intracavity power. The black dash line represents the upper bound of the resonance shift in devices with unnoticeable PR shift. The upper panels describe the fabrication sequence starting from two separate air-cladded chips and the symbols used in the corresponding data traces in the lower panel.}
    \label{summary}
\end{figure}

\section{Evolution of the PR effect during the fabrication process}

To further examine the impact of dielectric cladding on thin-film LN, we monitor the evolution of the PR effect during the fabrication flow. Before coated with cladding material, the as-fabricated air-cladded devices are first annealed at \SI{500}{\degreeCelsius} in a nitrogen environment for \SI{1}{\hour} to minimize the PR effect. The quality factor of all air-cladded devices is approximately $1.0 \times 10^6$.

The annealed air-cladded devices are processed in the following sequence: \SI{600}{\nano\meter} PECVD oxide coating at \SI{400}{\degreeCelsius}, thermal annealing, cladding layer removal, and re-annealing without cladding. After each step, the transmission of the devices is measured at different powers under a fixed scan rate of \SI{0.5}{\nano\metre/\second}. We fit the measured spectra using the model in section\,\ref{sec:scandynamic} and obtain the maximal resonance frequency shift to quantitatively characterize the strength of PR effect, as summarized in Figure \ref{summary}. Here, interactivity power is calculated to rule out the influence of quality factor and extinction among different resonances. Since the deformation of resonance shape is not observed in air-cladded devices, we use the resonance linewidth as an upper bound of resonant frequency shift and plot it in black dashed line in the figure.

Resonance deformation is observed directly after oxide coating, indicating the intensified PR effect after the PECVD process. And after the annealing, the PR effect is slightly enhanced compared to the unannealed case. We then remove the dielectric layer by BOE to check if the PR-free feature of the air-cladded devices could be restored. Unexpectedly, severe PR effect remains in these nominally uncladded devices. The resonance shifts are even larger under same intracavity power. Interestingly, the PR effect could be suppressed again after reannealing the uncladded devices at \SI{500}{\degreeCelsius} for \SI{1}{\hour} in nitrogen. No deformed resonances are observed in the re-annealed devices probed at similar power levels. The result suggest that fabrication process play a significant role in the activation and deactivation of the PR effect. The mechanism behind this could be the oxide coating/removal process modifies the interface state of LN, therefore aggravate the PR effect. The re-annealing at high temperature restores the interface properties and mitigate the PR effect. We notice that for all traces, the frequency shift does not increase linearly with the intracavity power. This confirms an optical power dependence of the space-charge generation process, which requires further investigation on the detailed mechanism of the PR effect in LN.

\section{Discussion and conclusion}

%, with a long time constant of ~ \SI{65.4}{min} after the optical illumination. This value is distinct from the relaxation time of PR effect in x-cut devices, which is typically on the order of \SI{10}{\milli\second}. 

Our result shows dielectric cladding layer and its related fabrication process could significantly impact the PR effect on LN. It is likely that interface states, which are introduced during the device processing, are responsible for the enhanced PR effect \cite{volk2008lithium,zhuang2012effects}. The interface states can provide both donor sites for photo-excitation and vacancy for excited carriers to occupy, thus influence the charge transportation process under light illumination \cite{hall1985photorefractive}. By removing the dielectric cladding and re-annealing, the interface traps are compensated and thus mitigate PR effect. Though we could remove the top cladding layer to mitigate the PR effect in LN microring resonator, the underneath LN-oxide bonding interface is beyond our control for unsuspended structure. Further investigation on suspended optical cavity could shed more light on the underlying mechanisms of the PR effect and its mitigation.

In summary, we present our findings on the PR effect on x-cut thin-film LN ring resonators. By removing the silicon dioxide cladding layer and applying proper heat treatment, we effectively suppress the resonance frequency shift in LN ring resonators induced by the PR effect. Although the underlying mechanism of our observation still requires further investigation, our work provides valuable guidelines for researchers battling with the instability of this powerful but sensitive material platform.

\section*{Funding}
This work is funded by ARO grant W911NF-18-1-0020. HXT acknowledges partial supports from NSF (EFMA-1640959) and the Packard Foundation. Funding for substrate materials used in this research was provided by DOE/BES grant under award number DE-SC0019406.  The authors thank M. Rooks, Y. Sun for assistance in device fabrication. 
\section*{Acknowledgments}
The authors thanks Michael Rooks, Yong Sun, Sean Rinehart and Kelly Woods for support in the cleanroom and assistance in device fabrication. 

\bibliography{reference}

\end{document}